\documentclass[sn-mathphys,Numbered]{sn-jnl}

\usepackage{graphicx}%
\usepackage{multirow}%
\usepackage{amsmath,amssymb,amsfonts}%
\usepackage{amsthm}%
\usepackage{mathrsfs}%
\usepackage[title]{appendix}%
\usepackage{xcolor}%
\usepackage{textcomp}%
\usepackage{manyfoot}%
\usepackage{booktabs}%
\usepackage{algorithm}%
\usepackage{algorithmicx}%
\usepackage{algpseudocode}%
\usepackage{listings}%

\def\4He{$^4$He}

\begin{document}

\title[Article Title]{Spherically symmetric counterflow turbulence in open geometry}

\author[1]{\fnm{Tom\'a\v{s}} \sur{Dunca}}
\author[1]{\fnm{Filip} \sur{Novotn\'{y}}}
\author[1]{\fnm{Marek} \sur{Tal\'{i}\v{r}}}
\author[1]{\fnm{Bal\'azs} \sur{Szalai}}
\author[1]{\fnm{Nikita} \sur{Ustinov}}
\author[1]{\fnm{Ladislav} \sur{Skrbek}}
\author*[1]{\fnm{Emil} \sur{Varga}}\email{emil.varga@matfyz.cuni.cz}

\affil*[1]{\orgdiv{Faculty of Mathematics and Physics}, \orgname{Charles University}, \orgaddress{\street{Ke Karlovu 3}, \city{Prague}, \postcode{121 16}, \country{Czech Republic}}}

\abstract
{We report preliminary results on spherical thermal counterflow  generated by a small central heater in an open geometry, an open bath of superfluid He~II, as closed-cell experiments could have introduced artifacts such as overheating and boundary-induced flows. In order to eliminate them, we measure second sound attenuation in a plane-parallel resonator. Our results are at variance with the previous experiments in closed spherical cavity that showed plateau in the steady-state vortex line density and its inverse time decay, neither of which is observed presently. We find that in open geometry the vortex line density $L$ increases steadily with counterflow velocity $v_\mathrm{ns}$, displaying a crossover between $L \propto v_\mathrm{ns}^2$ typical for
counterflow and $L \propto v_\mathrm{ns}^{3/2}$, characteristic for the quasi-classical scaling.
}


\maketitle

\section{Introduction}
\label{sec:intro}

Thermal counterflow of He~II  belongs to the class of most investigated quantum flows. In channels of constant cross-section it can be created by applying heat $\dot Q$ to the closed end of a channel with its other end open to the bath of He~II; $\dot Q$ is carried in a convective manner by the normal fluid of density $\rho_{\rm n}$. By conservation of mass, a superfluid current arises in the opposite direction and counterflow velocity is established: $v_\mathrm{ns} = \dot Q/ A \rho_{\rm s} \sigma T\,$, where $T$ is the temperature, $A$ is the channel cross-section and $\sigma$ stands for the specific entropy of He~II. Above $\approx 1\,$K, where He~II displays the two-fluid behavior, for small $\dot Q$, the flow of the normal viscous component is laminar and the flow of the superfluid component potential. Upon reaching a first critical velocity $v_\mathrm{ns}^{\rm{cr1}}$ a vortex tangle is created in the superfluid.  First detailed investigation of thermal counterflow was performed by Vinen \cite{VinenOlda,VinenOldb,VinenOldc,VinenOldd}, who introduced a phenomenological model describing a homogeneous random vortex tangle characterized by the vortex line density, denoted by $L$. On further increase of the  applied heat flux $\dot Q$, in dependence on geometry of the channel and interaction with its walls, normal fluid flow can remain laminar (so called T~I state \cite{Tough}) or become turbulent (so called T~II state \cite{Tough}). The turbulent He~II state represents a unique and very complex state of double turbulence \cite{PNASus,QTbook}.

In order to eliminate the influence of channel walls, one may consider spherical or cylindrical thermal counterflow, representing unbounded flows. It follows from the theoretical and numerical studies of spherical counterflow by Varga \cite{VargaSpherical} and by Inui \& Tsubota \cite{InuiTsubotaSpherical}, as well as from 2D studies of cylindrically symmetric flows \cite{BarenghiSergeevRickinsonCylinricalCF,SergeevBarenghiCylHVBK}, that a radial temperature gradient must exist in order for a stable steady-state counterflow to form. This theoretical prediction was confirmed in open He~II bath by the Prague group experimentally \cite{YunhuUs}. 

\begin{figure}
    \centering
    \includegraphics[width=0.9\linewidth]{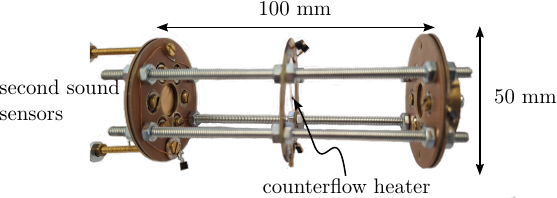}
    \caption{Photograph of the open acoustic resonator.}
    \label{fig:resonator}
\end{figure}

Subsequent Prague experiments \cite{SphericalCell} on spherical counterflow then displayed features which are in striking contrast with properties of channel counterflow. Specifically, the dependence of the averaged vortex line density versus power applied to the central heater displayed a temperature dependent plateau. These features were interpreted via turbulence being triggered in the normal fluid flow. Although, contradicting this, the temporal decay of vortex line density after the heat flux was switched off showed the $t^{-1}$ dependence at long times characteristic of Vinen turbulence, rather than the $t^{-3/2}$ that is typically present in late-stage decay of counterflow in the TII regime. Classical experiments demonstrate approximately Kolmogorov scaling also in spherically symmetric case where the turbulence is driven by vortex reconnections \cite{Matsuzawa}, therefore quasi-classical $t^{-3/2}$ decay would be expected.

The onset of this plateau was associated with a critical normal fluid Reynolds number, sometimes called Donnelly number $\mathrm{Dn}$, defined using the dynamic viscosity $\eta$ and normal fluid density $\rho_\mathrm{n}$ as ${\rm Dn}(r) = v_\mathrm{n} r \rho_{\rm n}/ \eta$. The onset of the plateau was found to occur at $\mathrm{Dn}\approx 50$, independent of temperature, with velocity $v_\mathrm{n}$ evaluated at the heater surface of radius $r= R_{\rm H}$.

However, the He~II sample inside the (almost) closed brass cell overheats, which was observed both in resonance frequency shift of the second sound mode and, at high enough $\dot Q$, by direct thermometry. Parasitic effects thus may have significantly affected the development of turbulence. For this reason, we decided to repeat this experiment in an open He~II bath 30 cm in diameter. Additionally, this experiment serves as a proof of concept for future, more detailed investigation of spherical counterflow under rotation - a model experiment for exploiting the condensed matter analogy with cosmological observations such as pulsar glitches and their decay \cite{Tsakadze,Bryn}.

\section{Open second sound resonator}

In order to eliminate overheating of the closed geometry and the resulting parasitic counterflow through the seams of the cell, we have designed and manufactured a plane-parallel second sound resonator, shown in Fig.~\ref{fig:resonator}. A ~1 x 1.5 mm oblate ellipsoid made from SMD resistor encased in
Stycast 2850FT is placed in its geometrical center and serves as the counterflow heater. The heater is soldered to two 50 \textmu m tinned copper wires which
split to current leads and voltage probes on the support structure. Metalized nuclepore membranes, 30 \textmu m thick with $\approx 100$~nm holes, are used as moving electrodes of capacitive transducer/receiver sensors of second sound.

\begin{figure}
    \centering
    \includegraphics[width=0.49\linewidth]{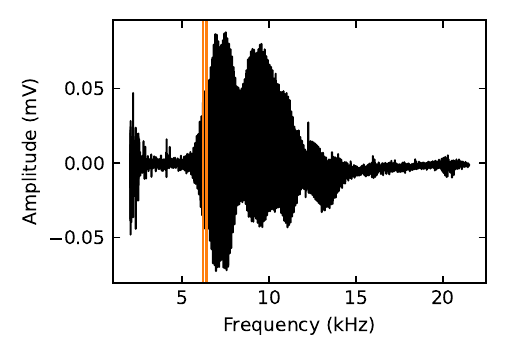}%
    \includegraphics[width=0.49\linewidth]{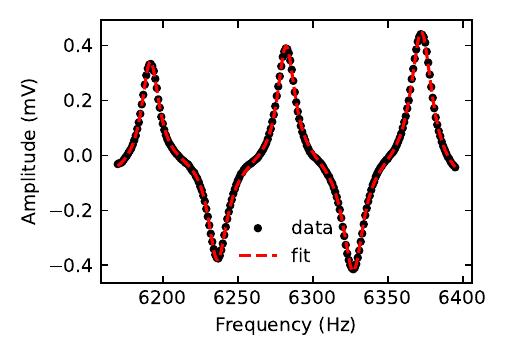}
    \caption{Left: Broad second sound spectrum at 1.3~K. Right: Narrow second sound spectrum between the lines indicated in the left panel and a fit to sum of 5 Lorentzian peaks.}
    \label{fig:ss-spectrum}
\end{figure}

A series of equidistantly spaced resonant modes, shown in the left panel of Fig.~\ref{fig:ss-spectrum} is observed in the frequency range 5 - 15 kHz with maximum quality factors exceeding 500. The turbulence-induced dissipation is measured using a section of the spectrum containing five resonant modes, shown in the right panel of Fig.~\ref{fig:ss-spectrum}. For quantitative description of vortex line density in the tangle generated by the thermal counterflow, these five resonances are fitted by five equidistantly spaced Lorentzians plus common linear background. It is known \cite{2ndSoundSpecIssueJLTP,Babuin} that for random, not extremely dense vortex tangle the second sound attenuation technique measures (if high $n\gg 1$ resonant harmonics are used) the average vortex line density in the volume between the second sound sensors
\begin{equation}
L=\frac{6 \pi \Delta_0}{\kappa B} \left (\frac{a_0}{a}-1 \right )\,,
\label{eq:VLD}
\end{equation}
\noindent
where $a_0$ and $a$ are, respectively, the attenuated and non-attenuated second sound amplitudes, $\Delta_0$ is the width of the non-attenuated Lorentzians, $\kappa \approx 10^{-7}$~m$^2$/s stands for the circulation quantum and $B$ denotes the dissipative mutual friction parameter, tabulated in \cite{DB}. Note that in this case the vortex line density is averaged over the volume of the acoustic mode. Assuming that the vortex tangle is localized close to the counterflow heater and does not extend past the resonator support structure, the shown vortex line densities below are rather a measure of \emph{total vortex length}.

\begin{figure}
    \centering
     \includegraphics[width=0.49\linewidth]{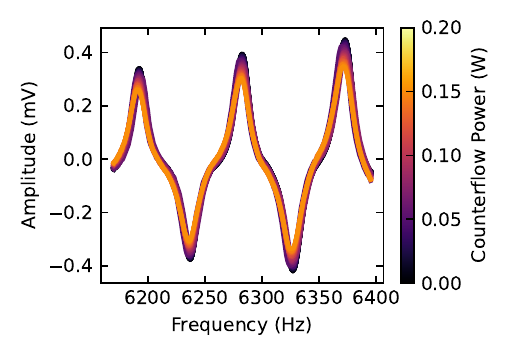}
    \includegraphics[width=0.49\linewidth]{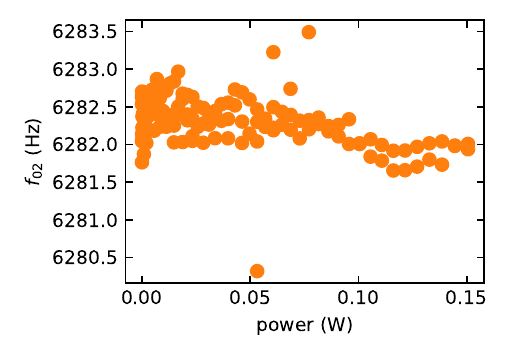}    
    \caption{Left: Attenuation of second sound resonances with increasing heater power. Shown data is for 1.3 K. Right: Shift in resonance frequency of one of the second sound resonances with increasing heater power at 1.3 K. }
    \label{fig:frequency-shift}
\end{figure}

\section{Results}
\label{sec:results}
\subsection{Steady-state spherical counterflow}

The steady counterflow velocity $v_\mathrm{ns}$ is parametrized by the radial counterflow velocity on a $r = 1$~mm shell around an idealized point source as
\begin{equation}
v_\mathrm{ns}(r)=\frac{1}{4 \pi r^2} \frac{\dot Q}{T \sigma \rho_{\rm s}}\,,
\label{eq:ss}
\end{equation}
\noindent
where $\sigma$ is the specific entropy and $\rho_{\rm s}$ is the density of the superfluid component of He~II. The left panel of Fig.~\ref{fig:frequency-shift} shows how the amplitudes of second sound peaks decrease with applied $\dot Q$. This is accompanied with the slight shift of the resonant frequency, as illustrated in the right panel of  Fig.~\ref{fig:frequency-shift}. Note that the maximum relative shift is for comparable heat input to the central heater about three orders of magnitude lower than that observed in the closed spherical cell \cite{SphericalCell}.

\begin{figure}
    \centering
    \includegraphics[width=0.89\linewidth]{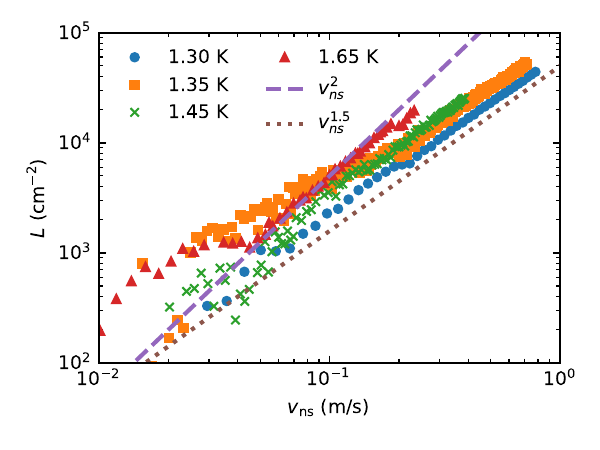}
    \caption{Scaling of vortex line density $L$ in statistically steady spherical counterflow with counterflow velocity $v_\mathrm{ns}$ for several temperatures.}
    \label{fig:L}
\end{figure}

Contrary to the previous experiments, we find that the vortex line density increases steadily with counterflow velocity, without any indication of a plateau at intermediate velocities. Moreover, Fig.~\ref{fig:L} shows that at the lowest attained temperatures we observe a clear departure from scaling of the form $L \propto v_\mathrm{ns}^2$, typical for thermal counterflow (also in cylindrical geometry \cite{SergeevBarenghiCylHVBK}), again in conflict with the results from a closed cell \cite{SphericalCell}. Scaling rather appears to follow the quasi-classical form $L \propto v_\mathrm{ns}^{1.5}$.

\subsection{Transient behavior}
Transient dynamics were obtained by measuring the second sound signal at a fixed frequency and switching the power between the spherical heater and a compensating heater placed in the bath far from the experiment, resulting in about 0.1~mK temperature stability of the bath. For temperatures other than 1.65 K, local overheating shifts the resonance by about 1~Hz (compared to about 20~Hz peak width), which was compensated for using support vector regression fit to the steady-state data.

\begin{figure}[h]
    \centering
    \includegraphics[width=0.49\linewidth]{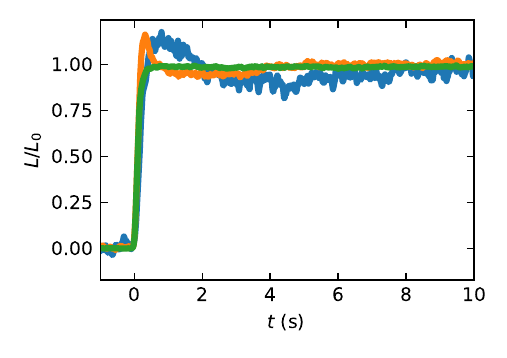}%
    \includegraphics[width=0.49\linewidth]{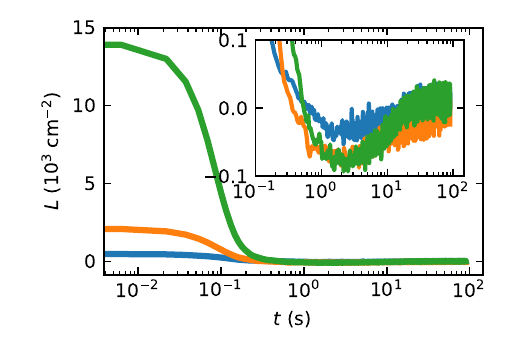}
     \includegraphics[width=0.49\linewidth]{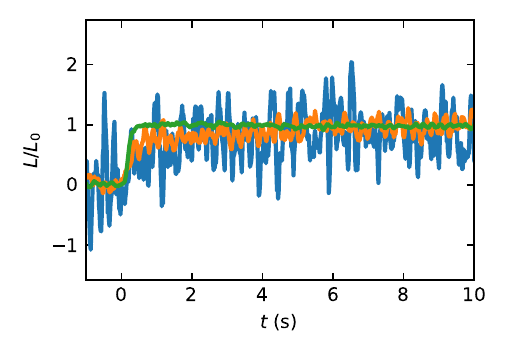}%
    \includegraphics[width=0.49\linewidth]{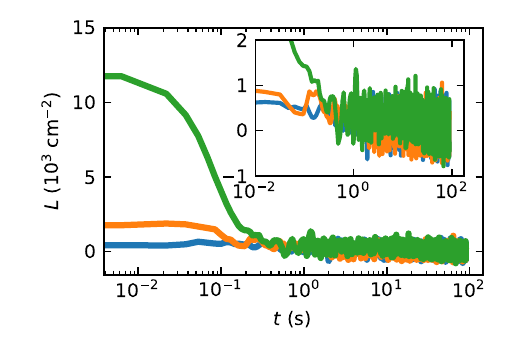}
    \caption{Vortex line density transients observed at 1.3~K (top) and  at 1.65~K (bottom). The left panels show the growth after heat flux is switched on, normalized to steady state $L_0$ and the right panels display the decay after heater is switched off. Curves from the same measurement are the same color in left and right panels. In all cases, $t=0$ corresponds to the instant when state of the heater was toggled. The insets in panels on the right show an enlarged view of the decay after the initial fast transient where a negative swing at intermediate times at 1.3~K is observed. The axes units in the inset are the same as in the main figure.}
    \label{fig:transient}
\end{figure}

Examples of the transient processes occurring at temperatures 1.3 and 1.65~K are shown in Fig~\ref{fig:transient}. In all cases, we observe a rapid transient between the attenuation levels. At 1.3~K, note the ``overshoot'' in the turbulence growth following switching on of the central heater and its relatively slow recovery toward the final equilibrium. This is somewhat reminiscent of the decay of spherical vortex shell observed in numerical simulations in Ref.~\cite{InuiTsubotaSpherical}. The data measured at 1.65~K display monotonic increase in the vortex line density without any overshoot, in contrast to 1.3~K data. We note in passing that similar ``overshoot'' was observed in earlier Prague experiments, studying channel counterflow generated both thermally \cite{LSCFB} as well as mechanically, by pressure gradient along a channel blocked by superleaks \cite{BabuinJLTP}.

The temporal decay, shown on the right in Fig.~\ref{fig:transient}, shows a rapid transient to a value of attenuation on the level of experimental noise without any long-term time dependence, contrary to channel counterflow and, indeed, to what was observed in the closed cell \cite{SphericalCell}. At lower temperatures, a "negative swing" in vortex line density is observed (i.e., the second sound attenuation is temporarily lower than for the reference state before the turbulence is switched on), which recovers approximately exponentially.

\section{Discussion}

The central heater, roughly shaped as an oblate ellipsoid 1.5~mm in main diameter, supplies to superfluid He~II bath rather high power up to about 300~mW. We therefore have to consider a possibility of cavitation or boiling in the vicinity of its surface. Based on direct comparison with the results of the group of Kono \cite{KonoJLTP,KonoPoster} we can exclude this possibility, as our data do not show any pronounced increase in second sound attenuation at the onset of boiling.

As for the steady-state spherical counterflow, contrary to the previous Prague experiments performed in a closed brass cell \cite{SphericalCell}, we find that the vortex line density increases steadily with counterflow velocity, without any indication of a plateau at intermediate velocities. Additionally, around $T = 1.65$~K the scaling seems to change from the form  $L \propto v_\mathrm{ns}^2$, typical for thermal counterflow to the quasi-classical form, $L \propto v_\mathrm{ns}^{3/2}$. Scaling also possibly changes with counterflow velocity, around $v_\mathrm{ns} \approx 0.1$~m/s, although this occurs only in a subset of studied temperatures. In analogy with the channel counterflow, one can calculate the normal fluid velocity at the surface of (spherical) central heater of the unit surface area as $v_{\rm n}^0 =\dot Q/(S_{\rm H}\rho \sigma T)$, where $S_{\rm H} \approx 4 \pi R_{\rm H}^2$ is the surface area of the heater. Assuming, as in Ref.~\cite{SphericalCell}, that the transition to turbulence of the normal fluid occurs near $\mathrm{Dn_{crit}} \approx 50$ results in counterflow velocity of order cm/s on the surface of the heater, which is on the edge of our resolution and about an order of magnitude below the possible transitions observed in the data. Therefore, the Donnelly number is unlikely to be a physically relevant parameter.

Disregarding the normal fluid turbulence, a naive application of the Vinen equation to the local counterflow velocity results in overall vortex line density that scales as $\dot Q^2$ regardless of the second sound detection mode geometry \cite{SphericalCell}. However, numerical simulations in the vicinity of solid spherical heater \cite{InuiTsubotaSpherical} suggest fast annihilation of the vortex lines closest to the heater (which was not observed in \cite{VargaSpherical}, likely due to the absence of solid walls near the origin). Assuming that the turbulence is continuously fed at a small scale $R_{\rm L}$ (e.g., the diameter of the heater
), we can estimate the size of this depletion region as the distance from the heater where the self-induced velocity of the smallest loops is overwhelmed by the converging superflow and vortices are thus ``sucked in'' to the heater. The radius $R_{\rm d}$ of the depletion region is given by the balance between the superfluid velocity and self-induced velocity of the small rings,
\begin{equation}
    \label{eq:small-loop-balance}
    \beta\frac{\kappa}{2\pi R_{\rm L}} = \frac{1}{4\pi R_{\rm d}^2}\frac{\rho_n}{\rho\rho_s} \frac{\dot Q}{\sigma T},
\end{equation}
where $\beta = \log(8R_{\rm L}/a) - 1/2$ is the logarithmic correction to the vortex ring velocity and $a\approx 10^{-10}\,$m is the vortex core size. From this it follows that $R_{\rm d} \propto \sqrt{\dot Q}$. If we now still retain the naive application of the Vinen equation to the local counterflow velocity (i.e., locally, $L = \gamma_{\rm V}^2v_\mathrm{ns}^2$ with $\gamma_{\rm V}$ the counterflow vortex line density parameter) outside of this depletion region, the total vortex line length is
\begin{equation}
    \label{eq:L-integral-32}
    \mathcal{L} = \int L\mathrm{d}V = \int_{R_{\rm d}}^\infty \gamma_V^2 v_\mathrm{ns}^2 4\pi r^2 \mathrm{d} r = \frac{\gamma_{\rm V}^2\dot Q^2}{4\pi (\rho_{\rm s} \sigma T)^2}\int_{R_{\rm d}}^\infty \frac{1}{r^2}\mathrm{d}r \propto \dot Q^{3/2},
\end{equation}
which agrees with the low-temperature behavior observed in the data. We stress, however, that the simple local application of the Vinen equation results in significant gradients of vortex line density near the heater, which will be exacerbated as temperature is increased. The local balance between vortex growth and annihilation assumed by the Vinen equation will have to be augmented by vortex diffusion and flux terms. This is beyond the scope of the current article and will be addressed in future studies.

We also note that a similar estimate can be made for the size of the turbulent blob localized around the heater: the size of the turbulent blob $R_\mathrm{blob}$ is such that the largest (and thus slowest) loops at this distance, which also have radius $R_\mathrm{blob}$ are not bound to the converging superflow, i.e.
\begin{equation}
    \label{eq:big-loop-balance}
    \beta\frac{\kappa}{2\pi R_\mathrm{blob}} = \frac{1}{4\pi R_\mathrm{blob}^2}\frac{\rho_{\rm n}}{\rho\rho_{\rm s}} \frac{\dot Q}{\sigma T}.
\end{equation}
This can be re-parametrized in terms of the counterflow velocity $v_\mathrm{ns}$ on the surface of the heater of radius $R_{\rm H}$ as $R_\mathrm{blob} = 2\pi r_{\rm H}^2\rho_{\rm n} v_\mathrm{ns} / (\beta\kappa\rho)$. Using values of Ref.~\cite{SphericalCell} at 1.35~K, the cell radius of $R_\mathrm{blob} \approx R_\mathrm{cell} \approx 10$~mm is reached for $v_\mathrm{ns} \approx 70$~mm/s, which is in the plateau region observed in \cite{SphericalCell}. This hints at a much simpler explanation of the plateau: the vortex line density stops increasing because the turbulent blob fills the cell and starts annihilating on the cell walls. The density starts growing again when the overheating of the cell reaches the point when the counterflow associated with the heat transfer across the cell walls and seams is sufficient to sustain turbulence.

The absence of the expected power law, or any extended-in-time, decay of vortex line density, however, remains a puzzle. One can imagine a degenerate form of turbulence consisting entirely of non-interacting vortex rings of radius $R_0$. These rings will annihilate in time $\tau = 2\rho_{\rm s}\pi R_0^2/(\gamma \beta)$ \cite{Barenghi1983}, where $\gamma \approx 5 \times 10^{-7}$~kg m$^{-2}$ s$^{-1}$ at 1.3~K is one of the formulations of the mutual friction parameters \cite{Barenghi1983}. Estimating the vortex ring size as $R_0^2 = L^{-1}$, vortex line density $L\approx 10^4$~cm$^{-2}$ ($\beta\approx 15$) would result in $\tau \approx 1$~s. Note that this vortex line density is the local density in the vicinity of the heater (for which an estimate of $10^4$~cm$^{-2}$ is not unreasonable), not the vortex line density averaged over the second sound mode volume shown in Figs.~\ref{fig:L} and \ref{fig:transient}.  While this provides a correct timescale for the sudden disappearance of the vortex tangle, it remains unclear why would the turbulence develop in this degenerate form.

\section{Conclusions}
We measured the steady-state and transient dynamics of vortex line density in spherical thermal counterflow in superfluid He~II in open geometry. Our preliminary results indicate a monotonic increase of vortex line density with counterflow velocity, at variance with previous experiments in a closed cell \cite{SphericalCell}. Moreover, we observe a change in the steady state scaling of vortex line density with counterflow velocity from $v_\mathrm{ns}^2$ at high temperatures to $v_\mathrm{ns}^{3/2}$ at low temperatures, which can be tentatively explained via vortex depletion region near the heater. Finally, we observe the transition dynamics (turbulence buildup and decay) to be rapid, without the hallmark power-law scaling of the decay. This decay occurs on timescale comparable with the decay of small vortex loops. Both the steady state scaling and decay highlight that a numerical simulation of spherical counterflow capable of reaching steady state is needed for full understanding. Further detailed experiments, including those under rotation utilizing the Prague rotating platform \cite{Dwivedi}, are under way.

\subsubsection*{Acknowledgments}
This research was supported by Czech Science Foundation (GA\v{C}R) under 25-16588S and 25-16386S. Fruitful discussions with Kimitoshi Kono are warmly acknowledged.

\section*{Declarations}
The authors have no competing interests to declare.

\section*{Data availability}
The data from Figs. 4 and 5 are available in a \href{https://doi.org/10.5281/zenodo.17144181}{Zenodo repository}

\bibliography{sn-bibliography}

\end{document}